\documentstyle[preprint,aps,epsf]{revtex}
\renewcommand{\vec}[1]{{\underline #1}}
\renewcommand{\Im}{\mbox{Im}}
\renewcommand{\Re}{\mbox{Re}}
\newcommand{\matr}[1]{{\underline{\underline #1}}}
\newcommand{\ccs}{$\rm CeCu_2Si_2$}
\newcommand{\lcs}{$\rm LaCu_2Si_2$}
\newcommand{\cerusi}{$\rm CeRu_2Si_2$}

\begin{document}

\title{The influence of crystal--field effects on the electronic transport
properties of heavy--fermion systems: a semiphenomenological approach}

\author{$^1$M.\ Huth$^*$ and $^2$F.\ B.\ Anders}

\address{$^1$Institut f\"ur Physik, Johannes Gutenberg--Universit\"at Mainz,
55099 Mainz, Germany\\ $^2$Department of Physics, University of
California, Davis, CA 95616-8677, USA\\ $^*$present address: Loomis
Laboratory of Physics, University of Illinois, Urbana--Champaign, IL
61801-3080, USA}

%%\date{\today}
\date{May 1, 1997}

\maketitle

\begin{abstract}
The electronic transport properties of heavy--fermion systems were
calculated based on a semiphenomenological approach to the lattice
non--crossing approximation in the limit of infinite local
correlations augmented by crystal--field effects. Within the scope of
this calculation using the linearized Boltzmann theory in the
relaxation time approximation the qualitative features of the
temperature--dependent resistivity, the magnetoresistivity and the
thermoelectric power can be successfully reproduced; this is
exemplified by a comparison with experimental results on \ccs\ .
\end{abstract}


\section{Introduction}
Strong electronic correlations are a crucial aspect of the physics of
heavy-fermion systems. The local Coulomb repulsion of the 4f or 5f
lattice sites causes the atomic part of the Hamiltonian to be of
non-bilinear form. Consequently the hybridization of the local
f-electrons with the itinerant states cannot be treated in a
conventional Feynman perturbation theory \cite{c1}. Several different
theoretical concepts have been developed during the last 10 to 15
years to treat this problem. These are rooted in the somewhat simpler
Kondo problem or
single--impurity problem that is now well--understood (for recent
reviews see \cite{c2}). If the f-moments are localized on a regular
lattice the problem is rendered even more difficult, since the
mechanism of coherent scattering for temperatures much below the Kondo
energy scale $T_K$ has to be introduced in the theory adequately in
order to describe the observed coherence effects in the electronic
transport properties. The first fully self-consistent microscopic
theory developed in this context is based on the lattice version of
the Non--Crossing Approximation (NCA) \cite{c3} and is therefore
called Lattice--NCA (LNCA) \cite{c4}. The LNCA is able to reproduce
the basic aspects of coherent scattering at low temperatures, such as
a quadratic dependence of the resistivity on temperature. Recently the
NCA was extended by a systematic continuation of the perturbation
expansion giving rise to an improved starting point for the lattice
approximation in the LNCA scheme \cite{c5}.

Considering the complicated nature of the LNCA and its numerical
elaboration a simplified approach including the essential aspects of
the Kondo lattice seemed to be helpful. For that reason a
semiphenomenological description of crystal field effects in Kondo
lattices based on the LNCA was developed which is
also applicable to the dynamical mean field theory (for reviews see
\cite{infiniteD}). This approach is able to
give further insight in the respective influences of the different
relevant energy scales in heavy--fermion systems on the measured
electronic transport quantities.

To begin with, the LNCA is briefly reviewed starting with the
formulation of the Anderson--lattice Hamiltonian in a generalized form
based on the irreducible representations of the point group of the
crystal. The results are exemplified for a Kramers ion in a tetragonal
crystal field. The derivation of the conduction electron Green
function is based on a semiphenomenological approximation to the local
excitation spectrum. Using the relation between the conduction
electron Green function and the transport relaxation time the
electronic transport properties are calculated within the linearized
Boltzmann theory. Finally, the calculations are complemented by a
comparison with experimental results.

\section{Description of the model}
The concept of the approximate treatment of the Anderson lattice for
the transport properties was introduced by Cox and Grewe
\cite{CoxGre88} and extended to  magnetotransport by Lorek, Anders and
Grewe \cite{c6}. It is based on the local approximation and uses the
local T-matrix calculated with the NCA/LNCA describing a single
scattering event of a conduction electron off the $f$-electrons. In
the next section this approximate treatment will be briefly reviewed
and completed by a consideration of the crystal field level scheme.
The transport coefficients are calculated using the transport integral
method of the linearized Boltzmann theory.

\subsection{The LNCA including the crystal field}
The starting point of our derivation is given by the Anderson--lattice
Hamiltonian. A spin--degenerate conduction electron
band couples to localized crystal field split ionic states by means of
hybridization matrix elements. Fluctuations on the ionic sites are
limited to those between singly occupied or empty states, i.\ e.\ the
repulsive interaction on the ionic sites is assumed to be infinite.
The Hamiltonian is given by the following expression
\begin{eqnarray}
  {\cal H} & = & \sum_{\vec{k}\sigma}
    \epsilon_{\vec{k}\sigma}c_{\vec{k}\sigma}^+c_{\vec{k}\sigma} +
  \sum_{\nu, \Gamma\alpha}
    E_{\Gamma\alpha}X_{\Gamma\alpha,\Gamma\alpha}^{\nu} +
\nonumber\\
   & & \sum_{\nu, \vec{k}\sigma, \Gamma \alpha }\left[
     V_{0, \Gamma \alpha }(\vec{k}\sigma)e^{i\vec{k}\cdot \vec{R}_{\nu}}
     c_{\vec{k}\sigma}^+X_{0, \Gamma \alpha }^{\nu} + \mbox{h.\ c.\ } \right] \!
 .
\end{eqnarray}
The first part describes the kinetic energy of the uncorrelated band
electrons. The energy of occupied local states is given by the second part.
Finally, the hybridization between local and itinerant states is formulated
by the third part of the Hamiltonian. $\Gamma$ denotes an irreducible
representation of the point group of the crystal, $\alpha$ a state of that
representation and $X_{\Gamma'\alpha',\Gamma\alpha}^{\nu}=
|\Gamma'\alpha'^{\nu}><\Gamma\alpha ^{\nu}|$ the Hubbard projection
operator at the site $\nu$.

In the periodic Anderson model the T--matrix $T_{\vec{k}\sigma}(z)$ is
related to the local Green function $F_{\Gamma \alpha ,\Gamma\alpha}(\vec{k},z)$
in the following way \cite{c7}
\begin{equation}
  T_{\vec{k}\sigma}(z) = \sum_{\Gamma'\alpha', \Gamma\alpha }
  V_{0, \Gamma\alpha}(\vec{k}\sigma) F_{\Gamma'\alpha', \Gamma\alpha}(\vec{k},z)
  V^*_{\Gamma'\alpha',0}(\vec{k}\sigma)
\end{equation}
which can be written in a more compact form by using a matrix formalism
\begin{equation}
  T_{\vec{k}\sigma}(z) =
    \vec{V}^t(\vec{k}\sigma)\matr{F}(\vec{k},z)\vec{V}(\vec{k}\sigma)
\end{equation}
introducing the hybridization vector $\vec{V}^t(\vec{k}\sigma)$ and the
Green function matrix $\matr{F}(\vec{k},z)$
\begin{equation}
  \vec{V}^t(\vec{k}\sigma) \equiv (V^*_{\Gamma_1\alpha_1,0}(\vec{k}\sigma),
  V^*_{\Gamma_1\alpha_2,0}(\vec{k}\sigma),
  \dots, V^*_{\Gamma_n\alpha_n,0}(\vec{k}\sigma))
\end{equation}
\begin{equation}
\left.  \matr{F}(\vec{k},z) \right|_{\Gamma'\alpha',\Gamma \alpha } \equiv
F_{\Gamma'\alpha',\Gamma \alpha }
(\vec{k},z) \equiv
\frac{1}{N}\sum_{\nu} e^{i\vec{k}\,\vec{R_{\nu}} }
  \ll X_{0, \Gamma'\alpha'}^{\nu}(\tau)X_{\Gamma \alpha ,0}^{0} \gg  \! .
\end{equation}
($N$: number of lattice sites)\\ With the exact equation of motion
\cite{c8} for the conduction electron Green function
\begin{equation}
  G_{\vec{k}\sigma}(z) =  G_{\vec{k}\sigma}^{(0)} +
  G_{\vec{k}\sigma}^{(0)}(z)T_{\vec{k}\sigma}(z)G_{\vec{k}\sigma}^{(0)}(z)
  \equiv \frac{1}{z-\epsilon_{\vec{k}\sigma}-\Sigma_{\vec{k}\sigma}(z)}
\end{equation}
we obtain the following expression for the band self--energy \cite{c7}:
\begin{eqnarray}
  \Sigma_{\vec{k}\sigma}(z) & = &
    \frac{\vec{V}^t(\vec{k}\sigma)\matr{F}(\vec{k},z)\vec{V}(\vec{k}\sigma)}
         {1+G_{\vec{k}\sigma}^{(0)}
    \vec{V}^t(\vec{k}\sigma)\matr{F}(\vec{k},z)\vec{V}(\vec{k}\sigma)}\nonumber\\
                            & = &
    \vec{V}^t(\vec{k}\sigma)\matr{F}(\vec{k},z)\vec{V}(\vec{k}\sigma)
    \sum_{\ell=0}^{\infty}
    (-1)^{\ell}\left[G_{\vec{k}\sigma}^{(0)}
    \vec{V}^t(\vec{k}\sigma)\matr{F}(\vec{k},z)\vec{V}(\vec{k}\sigma)\right]^{\ell}
\nonumber\\
                            & = &
    \vec{V}^t(\vec{k}\sigma)\matr{F}(\vec{k},z)\sum_{\ell=0}^{\infty}(-1)^{\ell}
    \left[\underbrace{
          \vec{V}(\vec{k}\sigma)G_{\vec{k}\sigma}^{(0)}
          \vec{V}^t(\vec{k}\sigma)}_{\equiv \matr{\Lambda}(\vec{k}\sigma, z)}
    \matr{F}(\vec{k},z)\vec{V}(\vec{k}\sigma)\right]^{\ell} \nonumber\\
                            & = &
    \vec{V}^t(\vec{k}\sigma)\matr{F}(\vec{k},z)
    \frac{1}{\matr{1}+\matr{\Lambda}(\vec{k}\sigma, z)\matr{F}(\vec{k},z)}
    \vec{V}(\vec{k}\sigma)
  \label{eq7}
\end{eqnarray}

The aim of the LNCA is to obtain an approximate expression for the
T--matrix and local Green function \cite{c4}. Unlike the NCA, in the
lattice the formal delocalization of a local f--electron via
hybridization into a band electron state may be followed by additional
scattering events on different lattice sites before the
re--localization into any local state takes place. In order to allow
for these scattering processes the concept of the effective lattice
site is introduced (all quantities refering to it are indicated with a
tilde). The scattering is subdivided into local and nonlocal events,
with the hybridization events starting and ending at a local lattice
site, respectively. The summation of all pseudo--local scattering
events results in the scattering matrix
$\vec{V}^t(\vec{k}\sigma)\matr{{\tilde{F}}}(z)\vec{V}(\vec{k}\sigma)$
of an effective site. Accounting for the exclusive--site condition
only for subsequent hybridization events the effective site Green
function $\matr{{\tilde{F}}}$ determines the local Green function
according to \cite{c4}
\begin{equation}
  \matr{F}(\vec{k},z) = \frac{1}{\matr{{\tilde{F}}}(z)^{-1}-
                        (\sum_{\sigma}\matr{\Lambda}(\vec{k}\sigma, z)
                                         -\matr{\Lambda}(z))}
  \label{eq8}
\end{equation}
with the definition
\begin{equation}
  \matr{\Lambda}(z) \equiv \frac{1}{\#k}
  \sum_{\vec{k}\sigma}\matr{\Lambda}(\vec{k}\sigma, z)
\end{equation}
Inserting equation (\ref{eq8}) in equation (\ref{eq7}) yields the
following expression for the band self--energy that is central for the
discussion of the transport properties within the scope of the
linearized Boltzmann transport theory:
\begin{equation}
  \Sigma_{\vec{k}\sigma}(z) = \vec{V}^t(\vec{k}\sigma)
    \frac{1}{\matr{{\tilde{F}}}(z)^{-1}+\matr{\Lambda}(z)-\matr{\Lambda}(\vec{k}
-\sigma, z)}
    \vec{V}(\vec{k}\sigma)
  \label{eq9}
\end{equation}
Note the fact that even a local approximation like the LNCA can include
the leading  anisotropies in the conduction electron self-energy  by
using an angular dependent hybridization vector
$\vec{V}(\vec{k}\sigma)$.

\subsection{Example: Three doublets; Kramers ion in tetragonal crystal field}
For further discussion we consider the level scheme of a Kramers ion
(e.\ g.\ Ce$^{3+}$) in a tetragonal crystal field. In this case the
Hund ground state multiplet $J=5/2$ splits into three magnetic
doublets. The hybridization is assumed to be spin--conserving without
$\vec{k}$--dependence. Since the local pseudo--spin is a good quantum
number, the hybridization couples only to one crystal field eigenstate
in either of the three doublets for a given band electron spin
direction. Consequently, the hybridization matrix is block diagonal
and the $6\times 6$--matrix decomposes into two $3\times 3$--matrices
with vanishing matrix elements in the upper/lower block matrix for
local down/up spin direction. Since the band electron Green function
has to be diagonal in the spin the same structure is given for the
matrix $\matr{\Lambda}(\vec{k}\sigma,z)$ resulting again in a block
diagonal form of $\matr{\Lambda}(z)$. Finally, the effective site
Green function matrix is diagonal due to the local hybridization. As a
consequence the band self--energy decomposes for a given band electron
spin into a sum of the self--energy contributions for the three
doublet states
\begin{equation}
  \Sigma_{\vec{k}\sigma}(z) = \sum_{j=1}^3 \frac{V_j^2\tilde{F}_j(z)}
                              {1+\tilde{F}_j(z)\Lambda_{jj}(z)} \! .
  \label{eq10}
\end{equation}

A crossover to the single impurity Kondo effect can be accomplished by
furnishing the scattering matrix with a prefactor $c_{imp}$ that represents
the concentration of the magnetic impurities
\begin{equation}
  V_j^2\tilde{F}_j(z) \quad \rightarrow \quad c_{imp} V_j^2\tilde{F}_j(z)
\end{equation}
After the expansion of the band self energy in equation (\ref{eq10})
with respect to the small quantity $c_{imp}$ the following result is
obtained

\begin{equation}
  \Sigma_{\vec{k}\sigma} (z) =
 c_{imp}  \left(
\sum_{j=1}^3 V_j^2\tilde{F}_j(z)
\right)
  \label{eq12}
\end{equation}

\subsection{Approximate treatment of the LNCA}
The local excitation spectrum is given by the spectral functions
$1/\pi\cdot \Im\tilde{F}_j(z)$. The dominant influence on the
transport properties is exerted by the many particle resonance
structure near the Fermi surface. In the following this resonance is
approximated by a Lorentzian curve
\begin{equation}
  \tilde{F}_j(\omega) = \frac{a(T/T_{Kj})}{\omega - \eta_j -i\gamma_j}
\end{equation}
which is in good correspondence to the results of a self--consistent
solution within the PNCA procedure \cite{c5}. The low energy scale
$T_{K0}$ determines both the width $\gamma = \pi/(2N+1)\cdot
k_BT_{K0}$  and the position $\eta = k_BT_{K0}$ of the main resonance.
The exact position of the resonance is difficult to specify accurately
since new results within the PNCA reveal a shift of the resonance
structure towards the Fermi energy in accordance with Friedel's sum
rule for a decreasing degeneracy $N_j$, unlike the former NCA result
\cite{c5}. The position and width of the $N_j$ degenerated excited
crystal field levels are given by $\Delta_{0j} + T_{K0}$ and $\gamma_j
= \pi/(2N_j+1)\cdot k_BT_{K_j}$ ($\Delta_{0j}$ crystal--field
splitting), respectively.

As is well--known from renormalization theory the low temperature
properties of a Kondo system are dominated by the low temperature
energy scale $T_K$. The argument of the temperature--dependent
function $a(T/T_{Kj})$ is therefore given by a relative temperature
for the respective crystal field levels. All calculations performed in
the following sections are based on the following temperature
dependence
\begin{equation}
  a(T/T_{Kj}) = \left\{\begin{array}{rcc}
                \frac{\gamma_i}{\Delta_j}(1-b(T/T_{Kj})^2) & : & T \ll T_{Kj}\\
                a_0(1-\frac{\ln{(T/T_{Kj})}}{\sqrt{(\ln{(T/T_{Kj})})^2+
                \pi^2S_j(S_j+1)}}) & : & T\geq T_{Kj}
                \end{array} \right.
  \label{eq14}
\end{equation}
The quadratic temperature dependence for $T \ll T_{Kj}$ is motivated
by the Fermi liquid character of the quasi-particle excitations at low
temperature. The prefactor $\gamma_i/\Delta_j$ ($\Delta_j = \pi
V_j^2N(\epsilon_F)$ Anderson width) is a consequence of the limiting
behavior of the local density of states for $T\rightarrow 0$ \cite{c9}
\begin{equation}
  N_j(\omega) = -\frac{1}{\pi}\Im{\tilde{F}_j(\omega -i\delta)} =
                   \frac{\gamma_j^2}{\pi\Delta_j[(\omega-\eta_j)^2+\gamma_j^2]}
\end{equation}
It reflects the many body nature of the resonance: only a fraction of
$\sum_j\gamma_j/\Delta_j < 1$ of $f$-electrons participate in the
scattering. In the temperature region $T\geq T_{Kj}$ we use the
results of the parquet diagram expansion of the Nagaoka--Suhl
equations for the sd--model \cite{c10}. The parameters $a_0$ and $b$
are fixed by adjusting equation (\ref{eq14}) to the temperature
dependence of the resonance height within the PNCA procedure
($N_j=2$). In the region between the low temperature and high
temperature limiting cases a polynomial fit based again on the PNCA
result \cite{c5} is used.

The influence of a magnetic field on the transport properties can be
accounted for by introducing a Kondo field $B_{Kj}$ that is given
by the following relation for the respective crystal field level
\begin{equation}
  k_BT_{Kj} = \mu_jB_{Kj}
\end{equation}
The argument of the function $a(x_j)$ is now given by a generalized form
that includes temperature and field on an equal base and guarantees an
isotropic field dependence in accordance with the assumed
$\vec{k}$--independent hybridization
\begin{equation}
  x_j =
\sqrt{\left(\frac{T}{T_{Kj}}\right)^2+\left(\frac{B}{B_{Kj}}\right)^2}
\label{scaling-x}
\end{equation}
Within the Lorentz approximation of the effective site Green function a
Zeeman splitting of the crystal field states has to be taken into account
\begin{equation}
  \tilde{F}_j(\omega) = \frac{a(x_j)}{\omega -(\eta_j+\mu_jB) -i\gamma_j}
\end{equation}

Nevertheless, in the paramagnetic phase of the model we expect that
there is only one low temperature energy scale, the lattice Kondo
temperature, which we set equal to $T_{K0}$. Equation
(\ref{scaling-x}) is only a natural way of parameterization of the
decrease of the quasi-particle spectral weight with temperature or
magnetic field and does not reflect a scaling law in the strict sense.
Additional low energy scales like the N\'eel temperature
\cite{GreweWel88} or the superconducting $T_c$ \cite{WelslauGre92} can
be calculated within the LNCA by analyzing the residual quasi-particle
interactions \cite{c4}.

\subsection{Relation to transport theory}
Based on the results of the preceding sections the band electron
Green--function is now used to give explicit expressions for different
electronic transport properties. Within the Green--function formalism
of the transport theory a comparison with the relaxation time
approximation of the linearized Boltzmann equation for the isotropic
(cubic) case gives the following expression for the relaxation time
\cite{c11}
\begin{equation}
  \tau_{\sigma}(z) = \int_{\mbox{band}} [\Im{G_{\sigma}(\omega,z)}]^2d\omega
\end{equation}
where the $\vec{k}$--dependence is averaged out.

This integral can be solved analytically for a symmetric energy band
of the width $2D$ yielding
\begin{eqnarray}
  \label{eq20}
  \tau_{\sigma}(z=x-i\delta) & = & \frac{\hbar}{\pi}\int_{-D}^D
  \left( \Im\frac{1}{x-i\delta-\omega-\Sigma_{\sigma}(x-
  i\delta)}\right)^2\,d\omega\nonumber\\
   & = & \frac{\hbar}{2\pi}
  \left\{\frac{\omega-(x-\Re \Sigma_{\sigma}(x-i\delta))}{[x-
  \Re \Sigma_{\sigma}(x-i\delta)-\omega]^2+[\Im \Sigma_{\sigma}(x-
  i\delta)]^2}\right. \\
   & + & \left.\left.\frac{1}{\Im \Sigma_{\sigma}(x-i\delta)}\arctan{\left[
  \frac{\omega-(x-\Re \Sigma_{\sigma}(x-i\delta))}{\Im
  \Sigma_{\sigma}(x-i\delta)}\right]}\right\}
  \right|_{-D}^{D}\nonumber
\end{eqnarray}
In the Fermi--liquid regime for $T\rightarrow 0$ the quasi-particle lifetime
at the Fermi energy diverges resulting in
$\Im\Sigma_{\sigma}(x-i\delta)\rightarrow 0$ and the second part in equation
(\ref{eq20}) dominates the relaxation time:
\begin{equation}
  \tau_{\sigma}(z=x-i\delta) \simeq \frac{\hbar}{2\Im \Sigma_{\sigma}(x-i\delta)
}
\end{equation}
Summing the spin--dependent relaxation times over the spin index leads to
the total transport relaxation time
\begin{equation}
  \tau(z=x-i\delta) = \sum_{\sigma} \tau_{\sigma}(z=x-i\delta) \! .
  \label{eq22}
\end{equation}
This can be used to determine several transport quantities like the
specific resistivity $\rho$, the thermopower $S$ and the Hall
coefficient $R_H$ based on the transport integrals $L_{mn}$ \cite{c12}:
\begin{eqnarray}
  L_{mn} & = & \int_{-\infty}^{\infty} \left(-\frac{\partial f}{\partial
  \omega}\right)\tau^m(\omega)(\hbar\omega)^n\,d\omega \\
  \rho & = & \frac{6\pi^2m^*}{e^2k_F^3}\frac{1}{L_{10}}\\
\label{ThermoPower}
  S & = & -\frac{1}{|e|T}\frac{L_{11}}{L_{10}}\\
\label{hall}
  R_H & = & -\frac{2}{ne}\frac{L_{20}}{L_{10}^2}
\end{eqnarray}

To conclude this section we give an explicit expression for the matrix
$\Lambda_{jj}(z)$ that is needed to calculate the band self--energy in
equation (\ref{eq10}). For a Gaussian density of states symmetric
around the Fermi energy $\rho(\epsilon) =
\rho(\epsilon_F)\exp{(-\omega^2/D^2)}$ the $\vec{k}$--sum in
$\Lambda_{jj}(z)$ can be performed as an energy integral
\begin{equation}
  \Lambda_{jj}(z) = \frac{1}{\#\vec{k}}\sum_{\vec{k}\sigma}
                    \frac{V_j^2}{z-\epsilon_{\vec{k}\sigma}} =
                    -i\pi V_j^2\rho(\epsilon_F)w(z/D)
\end{equation}
with the error function $w(z/D)$ \cite{c13}. Since the energy width of
the Kondo resonance is given by the small energy scale $k_BT_{Kj} \ll
D$ the approximation $w(x\ll 1) \simeq 1$ can safely be used; this
leads to
\begin{equation}
  \Lambda_{jj}(z) \simeq -i\pi V_j^2\rho(\epsilon_F) = -i\Delta_j
\end{equation}
Consequently, within the three doublet scenario, the calculation of
the transport quantities in the subsequent section is based upon the
following self--energy:
\begin{equation}
  \Sigma_{\sigma}(z=\omega\pm i\delta) = \sum_{j=1}^3 \frac{V_j^2
                 \tilde{F}_j(\omega\pm i\delta)}
                 {1 \mp i\tilde{F}_j(\omega\pm i\delta)\Delta_j}
  \label{eq29}
\end{equation}
In the low temperature region equation (\ref{eq29}) reflects
Matthiessen's law, according to which the resistivity is proportional
to the sum of the inverse relaxation times.

\section{Comparison with experimental results}
In the following we compare the results of the model calculation with
the transport properties of the heavy--fermion system {\ccs}. In
{\ccs} the crystal--field splitting between the low lying Kramer's
doublet and two energetically--degenerated excited doublets is much
larger than the low temperature energy scale $T_{K0}$ \cite{c14}.
Possible magnetic exchange scattering contributions to the resistivity
are neglected. Unlike the crystal--field splittings $\Delta_{0j}$ and
the Kondo temperature $T_{K0}$, the Kondo temperatures of the excited
doublets $T_{Kj}$ and the corresponding Anderson widths of the
respective levels $\Delta_j$ are not known. In order to derive
Anderson widths which are consistent with the fixed Kondo temperature
of the ground doublet and the Kondo temperatures of the excited
doublets the so called "poor man's" scaling result \cite{c2}
\begin{equation}
  k_BT_K = \sqrt{D \Delta} \exp{(-\pi|\epsilon_f|/2\Delta)}
  \label{eq30}
\end{equation}
is used with fixed values for the cutoff parameter $D=2.5\,$eV and the
position of the localized f--level $\epsilon_f = -2\,$eV (typical
values). For simplicity we assume equal Kondo--temperatures
$T_{K1}=T_{K2}$ for the energetically--degenerated excited two
doublets. The Kondo temperature for the ground doublet $|0>= -\eta|\pm
5/2>+\sqrt{1-\eta^2}|\mp 3/2>$ is set to 9\,K according to reference
\cite{c15} and $\eta = 0.467$ \cite{c14}.
The crystal--field splittings are known from inelastic neutron
scattering to be $\Delta_{01}=\Delta_{02}=350$\,K for both excited
levels $|1>= \sqrt{1-\eta^2}|\pm 5/2> + \eta|\mp 3/2>$ and $|2>=|\pm 1/2>$,
respectively. In cubic symmetry $\eta = \sqrt{1/6} = 0.408$ and $|1>$
and $|2>$ would form the $\Gamma_8$ quartet. Finally, the density of states 
at the Fermi 
level $N(\epsilon_F)$, the absolute value of Fermi wave vector $k_F$,
the effective mass of the non--correlated band electrons, and the
number density of charge carriers $n$ are chosen as 1/eV-states,
0.5/$\AA$, $8m_e$ ($m_e$: free--electron mass), and $2.5\cdot
10^{22}/cm^3$, respectively. The choice of these parameters is guided
by typical values for d--metals so as to give a good correspondence
with the measured resistivity. Additionally we neglected the
temperature dependency of the Lorentzian widths $\gamma_i$ in order to
maintain the lowest possible set of parameters. The maxima due to
higher crystal--field states will therefore appear more pronounced.
The following paragraph will show that the simplifying assumption of
an isotropic band structure results in quantitative differences
between the calculated and experimentally--determined higher--order
transport coefficients. Nevertheless, the aim of the present
calculation is to give a qualitative correspondence.

\begin{figure}[htb]
\unitlength1cm
\begin{picture} (18,6)
\put (0,0) {\epsfxsize=8cm\epsfbox{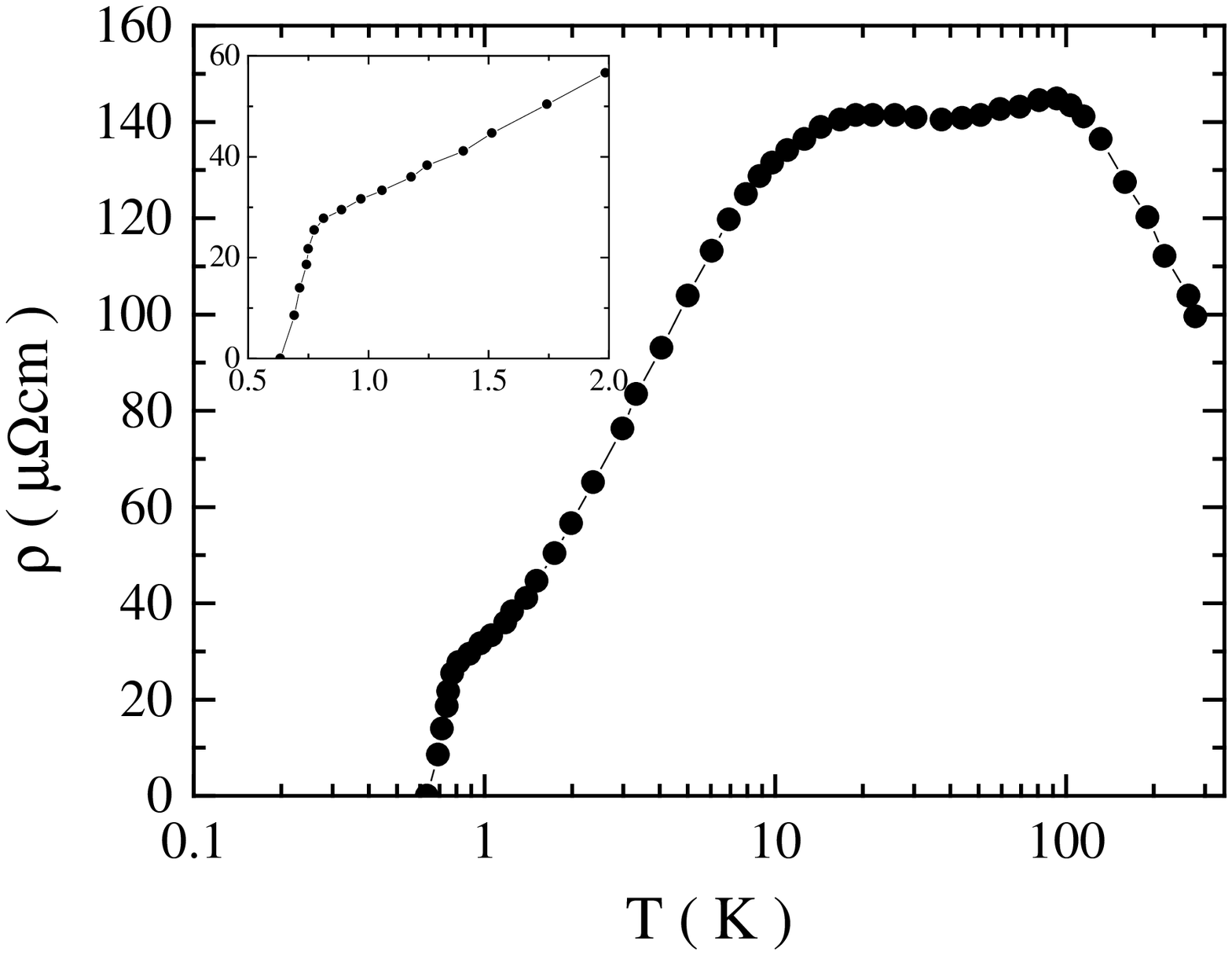}
\epsfxsize=8cm\epsfbox{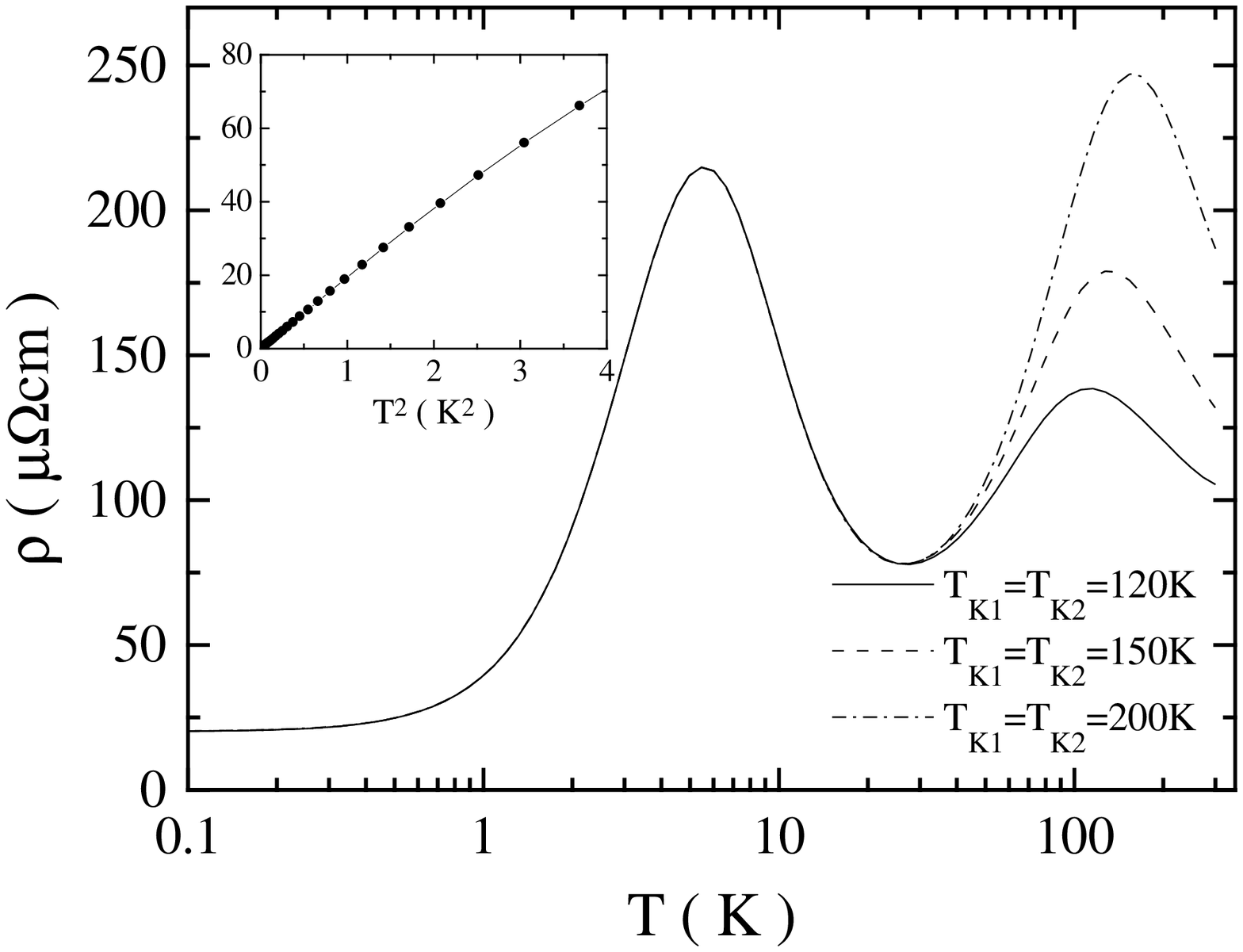} }
\end{picture}
\caption{Comparison of the temperature dependent resistivity of \ccs\ (left)
with the model calculation (right). The inset on the right shows the
calculated resistivity without residual resistivity and phonon
contribution as a function of $T^2$. The resistivity data is taken
from reference [\ref{c16}].}
\label{fig1}
\end{figure}

In figure \ref{fig1} a representative temperature--dependent
resistivity curve is contrasted with the model calculation for
different Kondo temperatures of the excited doublets. In order to
facilitate a comparison with the experimental data a residual
resistivity of 20\,$\mu\Omega$cm and the phononic resistivity
contribution of a \lcs\ reference--sample \cite{c16} was added to the
calculated curves.

The qualitative correspondence is best for an excited doublet Kondo
temperature of about 120\,K as far as the position of the high
temperature maximum in the resistivity is concerned. Interesting
enough, the width of the observed crystal--field transition by
Goremychkin et al.\ \cite{c14} is about 100\,K due to the strong
interactions of the f--electrons with the conduction band in good
correspondance with our estimate of the Kondo temperature of the
excited doublets.

\begin{figure}[t]
\unitlength1cm
\begin{picture} (18,5.8)
\put (-0.3,0) {\epsfxsize=8cm\epsfbox{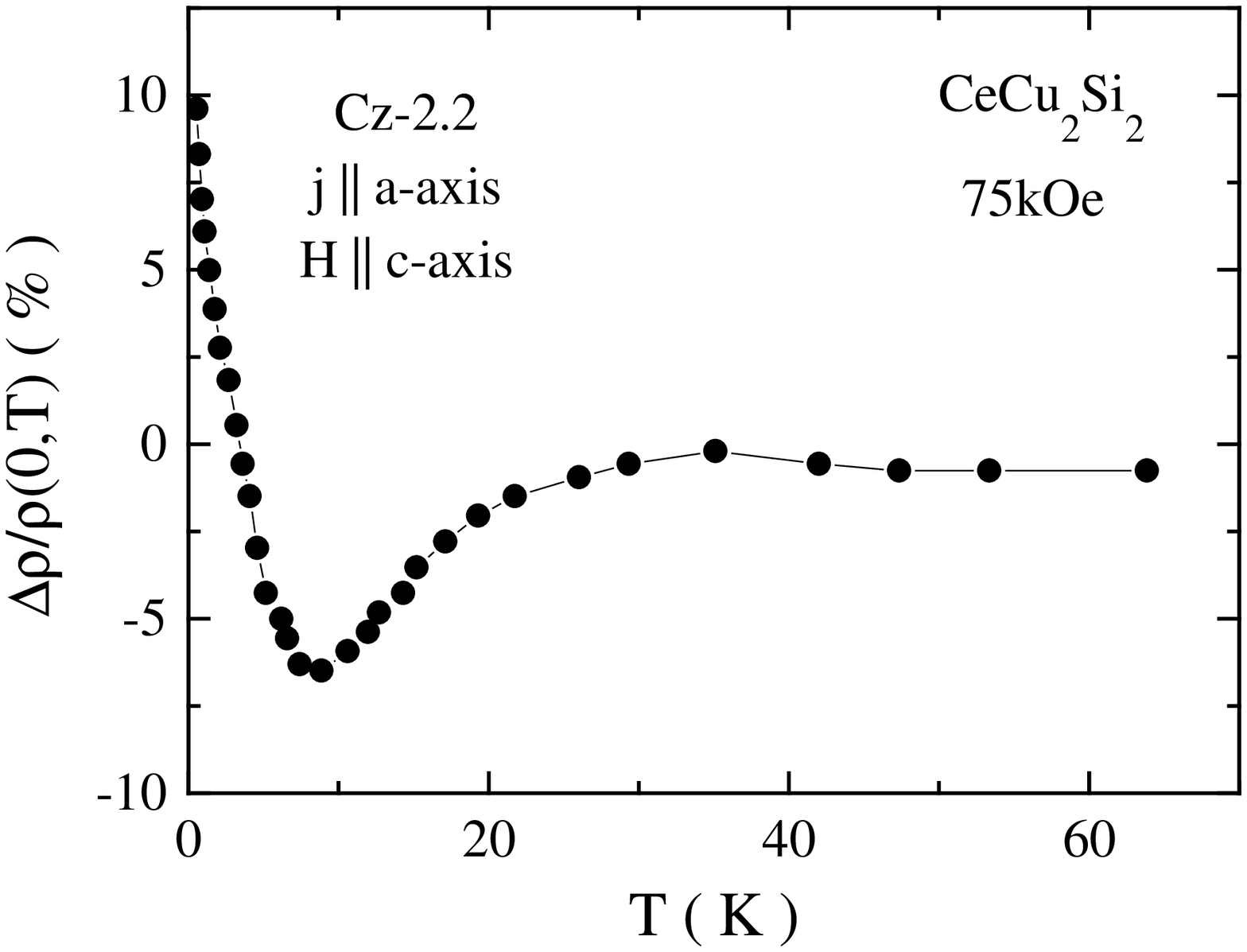}
\epsfxsize=8cm\epsfbox{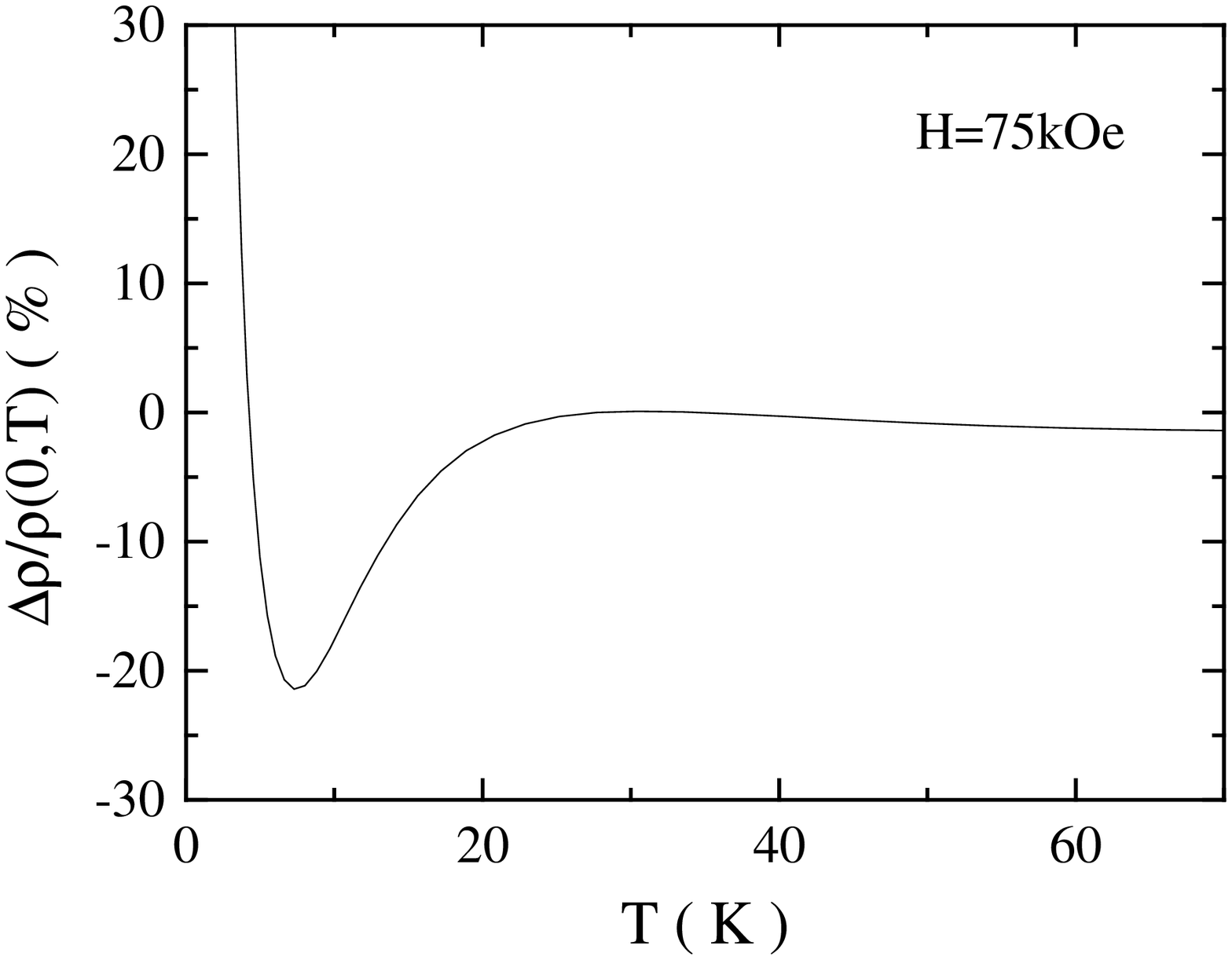} }
\end{picture}
\begin{picture} (18,5.8)
\put (0,0) {\epsfxsize=8cm\epsfbox{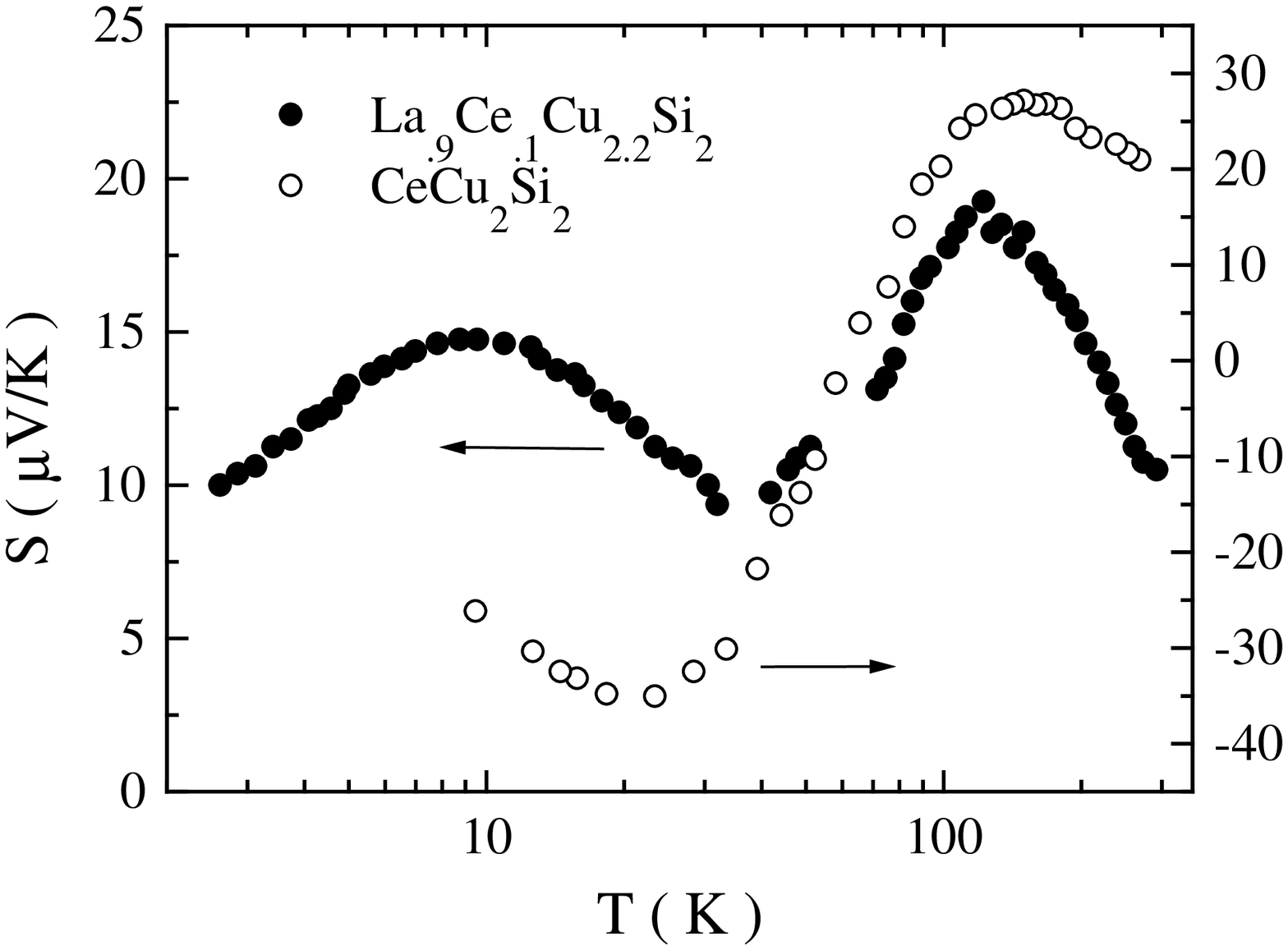}
\epsfxsize=8cm\epsfbox{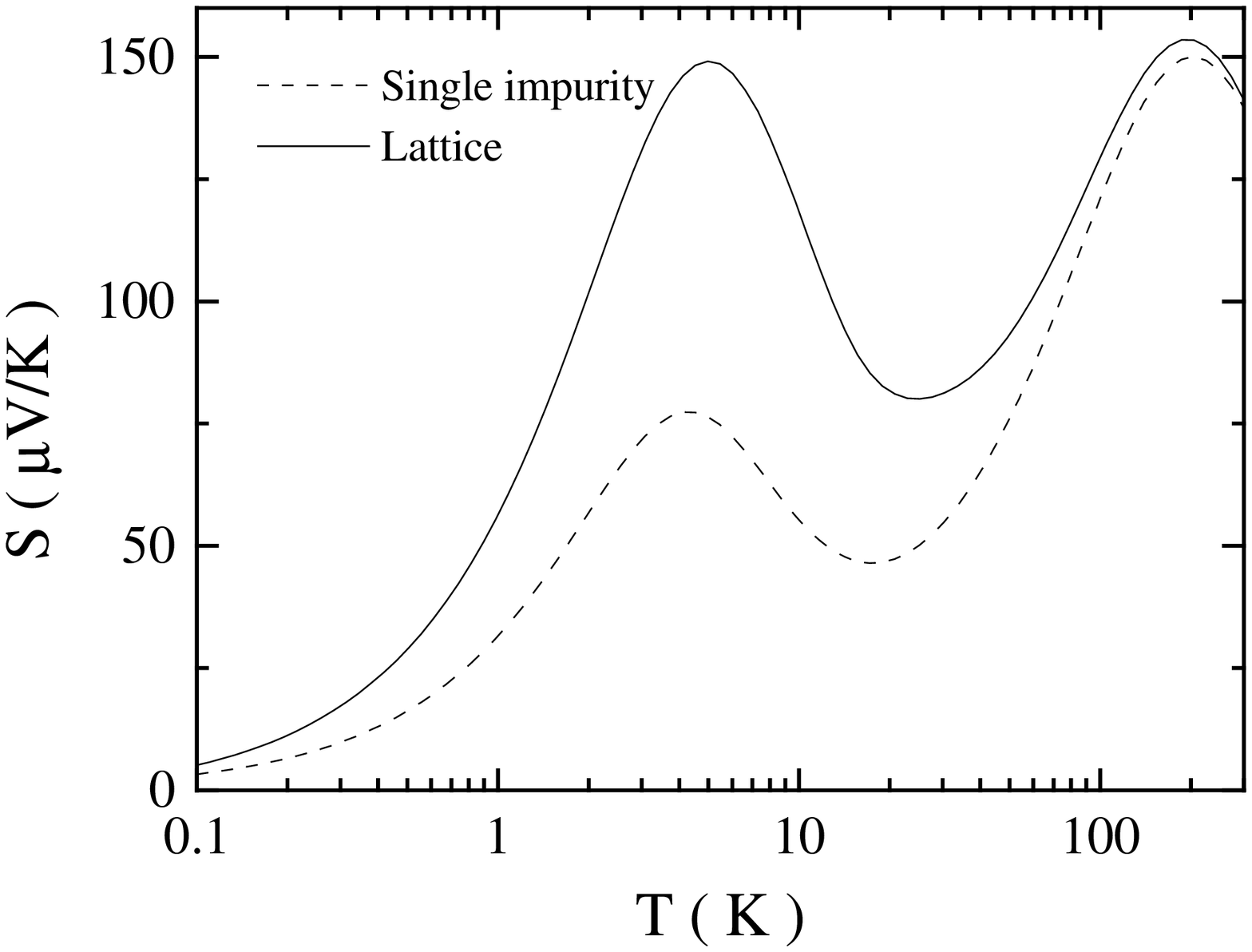} }
\end{picture}
\begin{picture} (18,5.8)
\put (0,0) {\epsfxsize=8cm\epsfbox{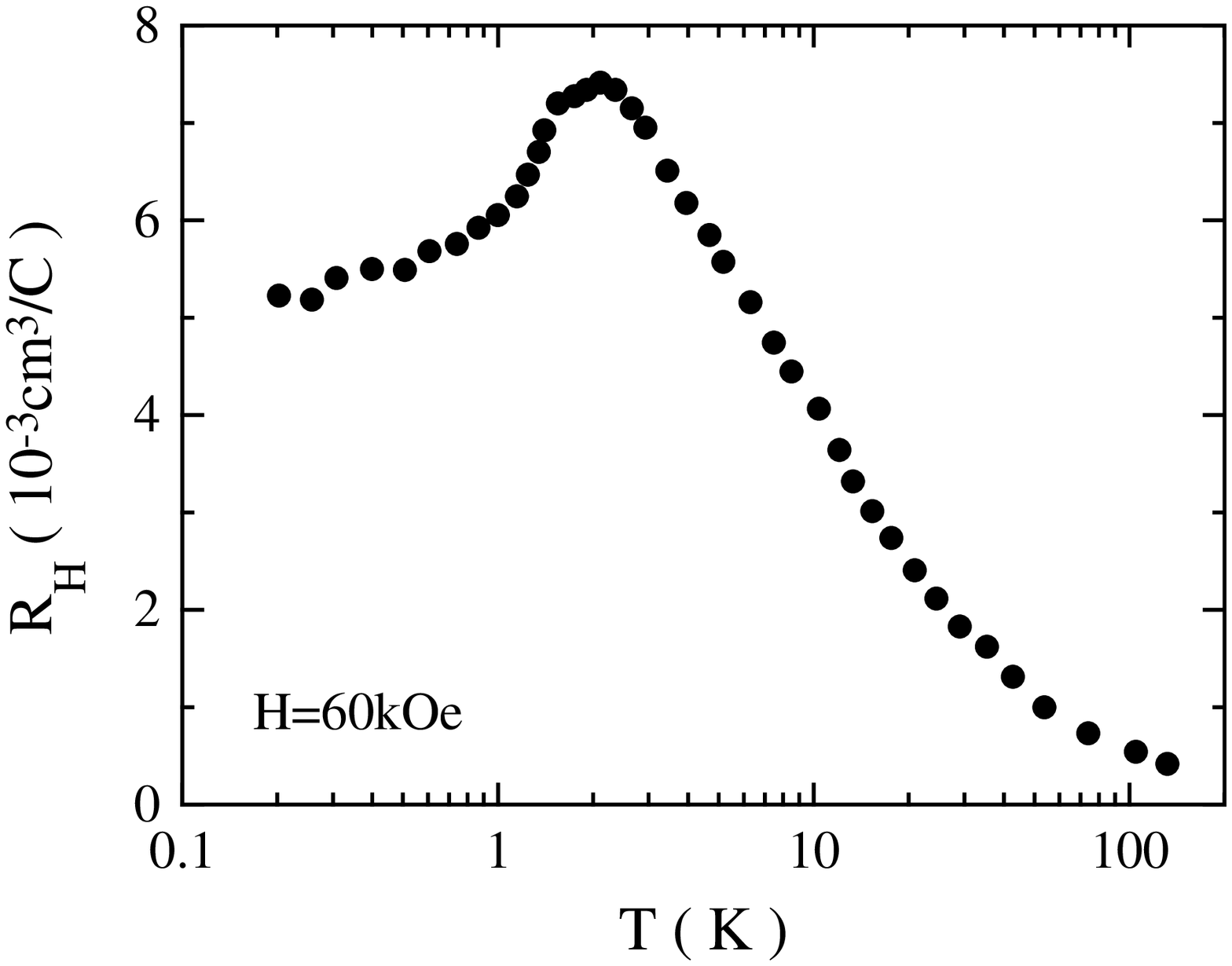}
\epsfxsize=8cm\epsfbox{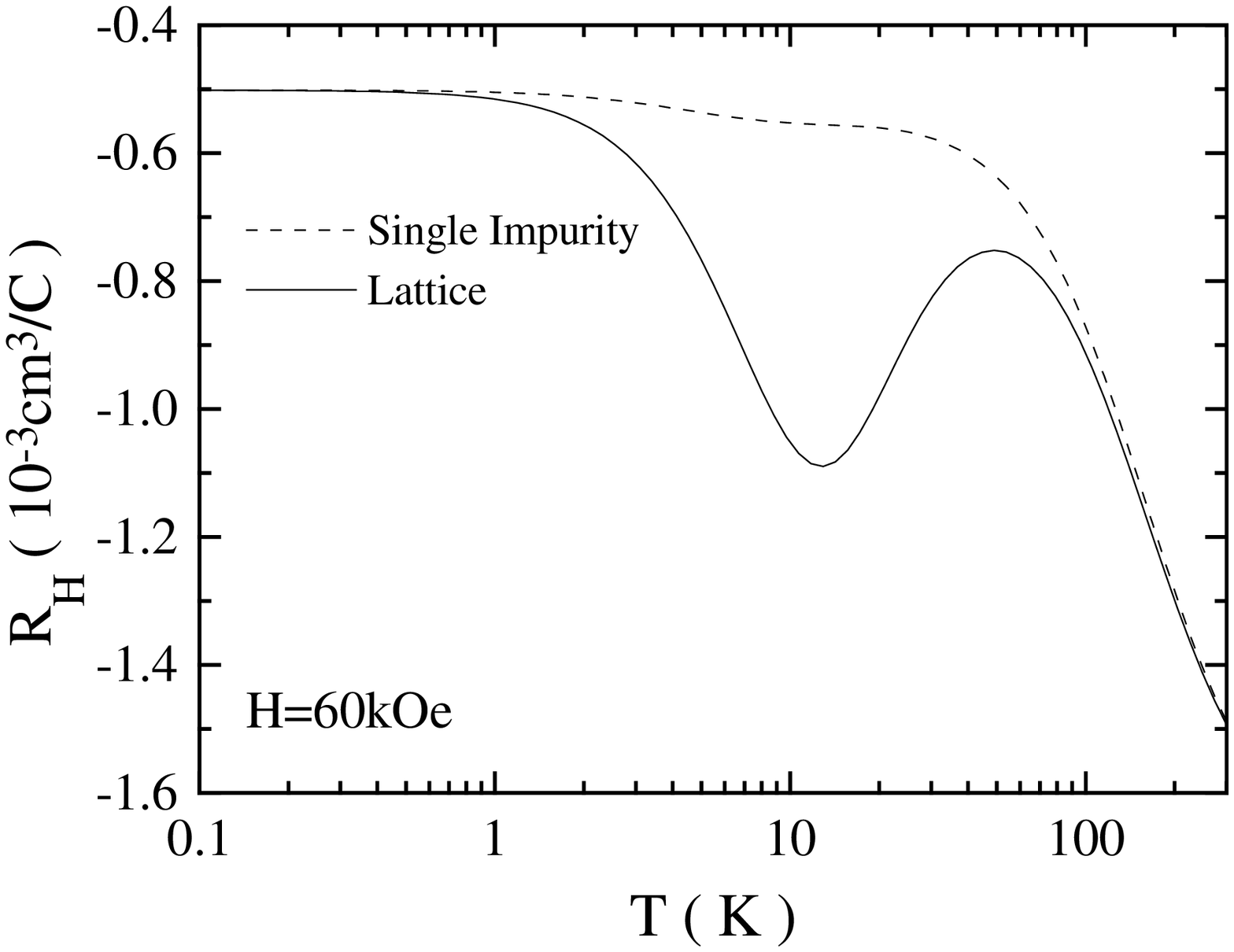} }
\end{picture}
\caption{Comparison of the temperature dependence of the magnetoresistivity
[\ref{c16}], the thermoelectric power[\ref{c17}]
 ({\ccs}) and [\ref{Steglich96}] ($\rm La_{0.9}Ce_{0.1}Cu_{2.2}Si_2$),
and the Hall coefficient [\ref{c18}] of \ccs\ (left) with the model
calculations
(right).}
\label{fig2}
\end{figure}

The low temperature maximum is more pronounced in the calculation.
This is also true for \ccs\ samples with lower residual resistivity.
The resistivity data of reference \cite{c16} was primarily taken since
this represents a common feature of disordered Kondo lattices. With
increasing disorder in the system a distribution of hybridization
strengths and positions of the local f--states is generated that cause
a much broader distribution of Kondo temperatures according to the
exponential dependence in equation (\ref{eq30}). A distribution of
Kondo temperatures might cause a significant rounding of the low
temperature maximum. Due to electronic correlations the
disorder--induced scattering rates grow faster than in uncorrelated
systems. As a consequence, the coherent scattering part is reduced and
the influence of disorder cannot be accounted for by a simple increase
in the residual resistivity \cite{c21}. As shown in the inset of
figure \ref{fig1} (right) the low temperature part of the calculated
resistivity follows a quadratic temperature dependence, as is expected
in the Fermi--liquid regime. The low temperature resistivity of
reference \cite{c16} reveals an interesting deviation from the
quadratic behavior following a linear dependence. This might be caused
by a a quantum critical point in \ccs\ as recently discussed by
Steglich et al.\ \cite{Steglich96}.

Based on the fixed parameter set with $T_{K1}=T_{K2}=120\,$K figure
\ref{fig2} shows a comparison of several additional transport quantities as
function of temperature. The left part represents experimental data from
various references whereas the right part shows the calculated quantities.

The calculated magnetoresistivity is in good correspondence with the
experimental data. Even the shallow maximum around 30\,K caused by the
excited crystal--field states is reproduced. Quantitatively the
magnetoresistivity at low temperatures reaches 800\% in the
calculation (with 20\,$\mu\Omega$cm residual resistivity) whereas the
experimental value is 10\,\% . The reason for the reduced
magnetoresistivity in the experimental data might also be traced back
to the same influence of disorder on the coherent scattering part in
the resistivity. Furthermore the calculation does not include possible
band--structure effects.

According to the calculation the Seebeck coefficient shows no
sign--reversal; this is in contradiction to the experimental result.
Furthermore the low temperature maximum does not appear in the
experimental data. Furthermore, the absolute values in the calculation
are sigificantly higher. This could be attributed to a
disorder--induced smoothening of the sharp features at the Fermi level
in the samples. Some general remarks seem to be appropriate here. The
temperature dependence of the thermopower of heavy--fermion systems
shows a wide variety of different features. In the trivalent cerium
systems a sign reversal with a negative low temperature minimum is a
common feature in concentrated systems. Recently Kim and Cox
\cite{KimCox96} discussed the possible realization of a two-channel
Kondo impurity model in Ce$^{3+}$ predicting a large negative
thermopower in a cubic crystal symmetry and non-Fermi-liquid
signatures in thermodynamic quantities. In dilute Cerium systems,
however, a negative thermopower has never been observed. In recent
measurements on La$_{0.9}$Ce$_{0.1}$Cu$_{2.2}$Si$_{2}$
\cite{Steglich96} a clearly positive thermopower is observed which is
in strikingly good qualitative agreement with our calculation (see
single--impurity calculation of $S(T)$ in figure \ref{fig2}). For
temperatures $T\ge T_K$ the lattice and the impurity thermopower
calculated within our model are very similar since coherence no longer
plays a role and the impurity concentration explicitly cancels in
equation \ref{ThermoPower}. On the other hand the thermopower
sensitively measures the asymmetry of the scattering rate above and
below the chemical potential. In our opinion the negative thermopower
of CeCu$_{2}$Si$_{2}$, which is still not properly understood, points
towards additional lattice effects not included in our model. Besides
a possible influence of the band structure non--local quasiparticle
interactions, which are not contained in the local approximation,
cause significant renomalizations of the one-particle properties; this
is part of the LNCA concept \cite{c4}. These interactions  mediate
short--range antiferromagnetic fluctuations whose correlation length
can grow with
decreasing temperature. Assuming a quantum critical point
($T_N\rightarrow 0$) the fluctuations can account for the
observed non-Fermi-liquid behavior in CeCu$_{2}$Si$_{2}$  above the
superconducting $T_c$ \cite{Steglich96}.

With increasing valence instability a crossover to an overall positive
thermopower with a second low temperature maximum is observed (see e.\ g.\
in \cerusi\ \cite{c22}). One possible reason for the negative component of
the thermopower is given by intersite spin--interactions. Consequently, due
to the onset of real charge fluctuations on the f--sites the negative part
of the thermopower is suppressed \cite{c23}. Spin--spin interactions are
not included in the calculation so the qualitative feature with two maxima
in the calculated thermopower might be more representative of systems like
\cerusi\ .

We conclude this section with some remarks concerning the Hall coefficient.
In this case the discrepancies between the calculation and the experimental
data are especially pronounced. \ccs\ shows a skew--scattering behavior
typical of all heavy--fermion systems and this is not reproduced by the
calculation. On the other hand, the Hall coefficient can vary appreciably
in magnitude and sign within a set of samples of the same heavy--fermion
system.  The intricate influence of inter--site spin interactions might be
responsible for the strong deviations of the calculations from the
experimental data. Nevertheless, the calculations predict in the
Fermi--liquid regime a negative quadratic temperature dependence. With
increasing temperature the skew--scattering part of the Hall coefficient
increases and eventually overcompensates the negative Fermi--liquid
contribution. As a result, a low--temperature minimum in the Hall
coefficient develops. This is experimentally observed in several
heavy--fermion systems and was discussed in more detail in \cite{HuthRH}.

\section{Conclusions}
The electronic transport properties of heavy--fermion systems were
calculated on the basis of a semiphenomenological approach to the
lattice non--crossing approximation in the limit of infinite local
Coulomb replusion  augmented by crystal--field effects. The calculation is
able to reproduce the qualitative features of the
temperature--dependent resistivity, the magnetoresistivity and the
thermoelectric power, as exemplified by a comparison with experimental
data of \ccs\ and  La$_{0.9}$Ce$_{0.1}$Cu$_{2.2}$Si$_{2}$. The
skew--scattering characteristic of the temperature dependent Hall
coefficient is not reproduced. Nevertheless, the lack of agreement
between the experimental data and the calculation could presumably be
traced back to secondary effects which are not included in our model.
The disorder--induced suppression of coherent scattering in the
resistivity and inter--site spin interactions that might especially
influence the thermoelectric power and the Hall coefficient are not
taken into account.

Within the scope of this approach an extension to include crystal--field
excitations to magnetic singlet states can be easily accomplished by adding
the respective resistivity or, more generally, the self--energy contributions
of the singlet states as described by Cornut and Coqblin \cite{c20}. This
permits the calculation of the transport properties of the uranium--based
heavy--fermion systems, which tend to be in a 5f$^2$--state \cite{c24}.

A more sophisticated transport theory based on the Kubo formalism
might be neccessary to account for the generally anisotropic band
structure and probably an anisotropic hybridization. This might result
in a more quantitative agreement between the experimental data and the
calculated transport coefficients. However, in our opinion the
consideration of non--local quasiparticle interactions is essential in
order to account for the observed behavior of the thermopower and Hall
coefficient. This is surely beyond the present approach and should be
devoted to calculations based on a fully microscopic description of
the electronic transport properties in heavy--fermion systems.

\acknowledgments
This work was supported by the Deutsche Forschungsgemeinschaft through SFB
252. Additionally one of us (FBA) was supported by the Deutsche
Forschungsgemeinschaft, in part by the National Science Foundation
under Grant No. PHY94-07194, and the US Department of Energy, Office of
Basic Energy Science, Division of Materials Research. FBA would like
to thank the Institute for Theoretical Physics (ITP) in Santa Barbara,
California, USA, for its hospitality where part of the work has been
performed.

\end{document}